\documentclass[pra,twocolumn,showpacs,preprintnumbers,amsmath,amssymb]{revtex4}

\usepackage{graphicx}% Include figure files
\usepackage{dcolumn}% Align table columns on decimal point
\usepackage{bm}% bold math

\usepackage[T1]{fontenc}
\usepackage[latin1]{inputenc}
\usepackage[english]{babel}
\usepackage{amsmath}
\usepackage{amssymb}
\usepackage{amstext}
\usepackage{bbold}
\usepackage{braket}
\usepackage{MnSymbol}
\usepackage{color}
\usepackage{relsize}
\usepackage{subfigure}
\usepackage{booktabs}
\usepackage{natbib}

\newcommand{\vs}[1]{\ensuremath{\boldsymbol{#1}}}
\newcommand{\vect}[1]{\ensuremath{\bm{#1}}}
\newcommand{\eq}[1]{\begin{equation} #1 \end{equation}}
\newcommand{\eqa}[1]{\begin{eqnarray} #1 \end{eqnarray}}

\newcommand{\be}{\begin{equation}}
\newcommand{\ee}{\end{equation}}
\newcommand{\ba}{\begin{eqnarray}}
\newcommand{\ea}{\end{eqnarray}}

\newcommand{\ue}{\mathrm{e}}

\begin{document}

\title{
Quantum disorder in the spatially completely anisotropic triangular lattice \newline
I: Heisenberg $S=1/2$ antiferromagnet
}
\author{Philipp Hauke}
    \email{philipp.hauke@icfo.es}
    \affiliation{ICFO -- Institut de Ci\`{e}ncies Fot\`{o}niques, Parc Mediterrani de la Tecnologia, Av.\ Carl Friedrich Gauss 3, 08860 Castelldefels, Spain}
    \affiliation{Institute for Quantum Optics and Quantum Information of the Austrian Academy of Sciences, 6020 Innsbruck, Austria}

\date{\today}

\begin{abstract}
Spin liquids occuring in 2D frustrated spin systems were initially assumed to appear at strongest frustration, but evidence grows that they more likely intervene at transitions between two different types of order.
To identify if this is more general, we here analyze a generalization of the spatially anisotropic triangular lattice (SATL) with antiferromagnetic Heisenberg interactions, the spatially \emph{completely} anisotropic triangular lattice (SCATL). Using Takahashi's modified spin-wave theory, complemented by exact diagonalizations, we find indications that indeed different kinds of order are always separated by disordered phases. Our results further suggest that two gapped non-magnetic phases, identified as distinct in the SATL, are actually continuously connected via the additional anisotropy of the SCATL. Finally, measurements on several materials found magnetic long-range order where calculations on the SATL predict disordered behavior. Our results suggest a simple explanation through the additional anisotropy of the SCATL, which locates the corresponding parameter values in ordered phases. The studied model might therefore not only yield fundamental insight into quantum disordered phases, but should also be relevant for experiments on the quest for spin liquids. 
\end{abstract}

%    75.10.Jm Quantized spin models
%    75.10.Kt Quantum spin liquids
%    75.30.Ds Spin waves
%    75.30.Kz Magnetic phase boundaries (including magnetic transitions, metamagnetism, etc.)
%    75.50.Ee Antiferromagnetics

\maketitle

\section{Introduction}

Understanding magnetically disordered quantum materials is of fundamental interest, e.g., for layered magnetic insulators/metals in which
magnetism is disrupted by charge doping, leading to dramatic phenomena such as high-temperature superconductivity \cite{Kastner1998,Lee2006,delaCruz2008}.
Also, disordered quantum phases can have excitations which are fractionalized even in two dimensions \cite{Anderson1973}.
However, classical order is typically quite resilient, especially in two or three dimensions \cite{Dyson1978,Kennedy1988,Manousakis1991,Misguich2004}. To disrupt the classical order and reach quantum-disordered phases like valence-bond solids or resonating valence-bond states, it is assumed that frustration could be a crucial ingredient \cite{Anderson1973,Fazekas1974}. 
Here, we want to investigate where such quantum-disordered phases appear in a two-dimensional, antiferromagnetic (AFM) Heisenberg model with highly tunable frustration.

\begin{figure}
	\centering
	\includegraphics[width=0.49\textwidth]{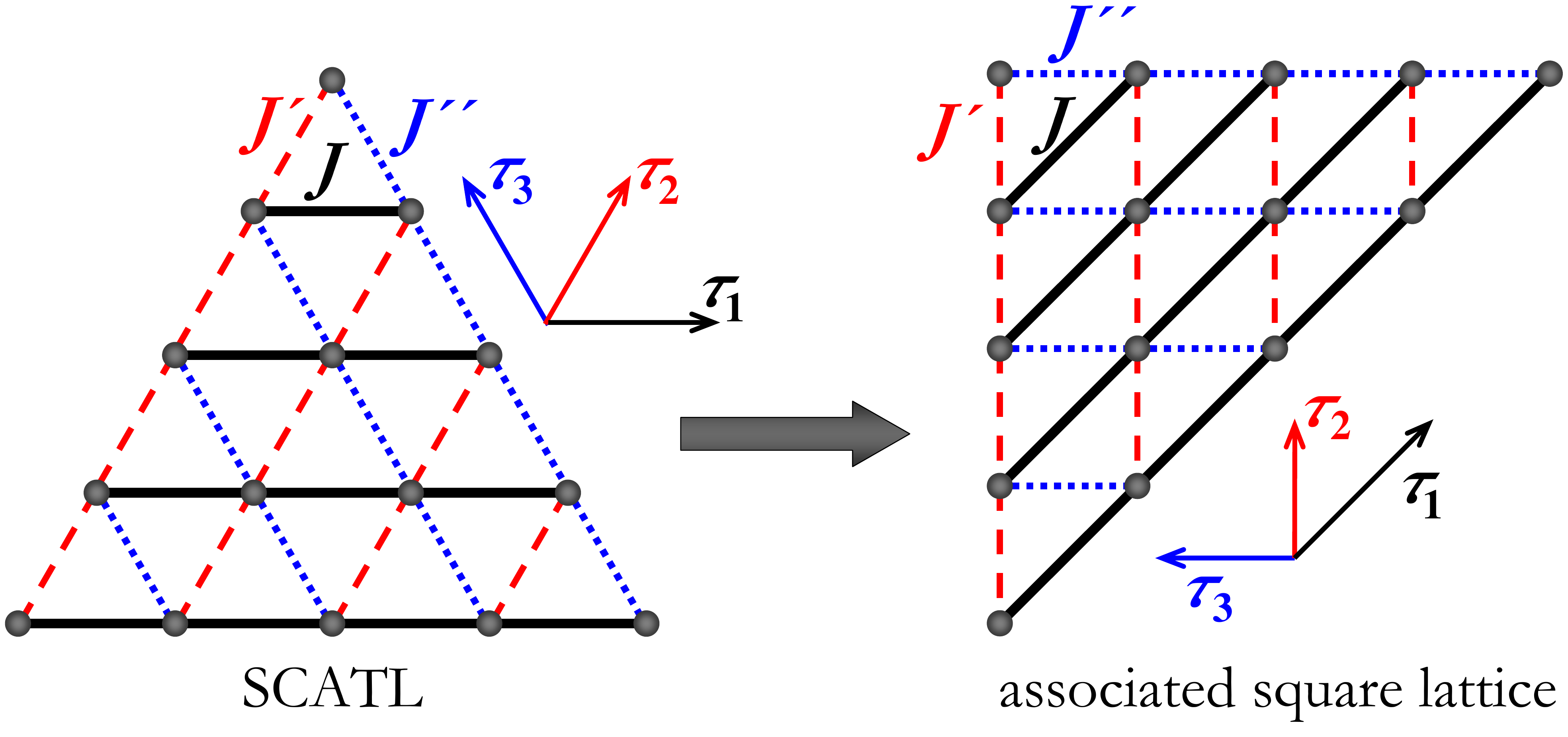}
	\caption{
	  {\bf Geometry of the SCATL.} The spins (gray bullets) are coupled to NNs along the lattice vectors $\vs{\tau}_{1,2,3}$ by the couplings $J_{\vs{\tau}_{1}}\equiv J$, $J_{\vs{\tau}_{2}}\equiv J'$, and $J_{\vs{\tau}_{3}}\equiv J''$, which can all be mutually different. 
	  The right part of the Figure shows the associated square lattice.
	  The shown geometry is the one used in the ED of the 15-site system, chosen for maximal symmetry between all three couplings. 
	  \label{fig:geometry}
	}
\end{figure}

The Heisenberg Hamiltonian describes a large variety of magnetic materials. It reads
\begin{equation}
 \label{eq:HS}
  H_{\text{S}}=
  \sum_{\braket{i,j}} J_{ij} ~{\bm S}_i\cdot{\bm S}_j,
\end{equation}
where ${\bm S}_i$ is a Heisenberg spin-$S$ operator at site $i$ (here we are interested in the extreme quantum limit $S=1/2$).
We consider a triangular geometry where the nearest-neighbor (NN) couplings $J_{ij}$ along all three lattice directions are different, the spatially completely anisotropic triangular lattice (SCATL), see left side of Fig.~\ref{fig:geometry}. 
For simplicity, we will work throughout this paper in the associated square lattice (right side of Fig.~\ref{fig:geometry}), where the vectors connecting NN sites are $\vs{\tau}_{1}\equiv(1,1)$, $\vs{\tau}_{2}\equiv(0,1)$, and $\vs{\tau}_{3}\equiv(-1,0)$, and define  $J_{\vs{\tau}_{1}}\equiv J$, $J_{\vs{\tau}_{2}}\equiv J'$, and $J_{\vs{\tau}_{3}}\equiv J''$. 
This model generalizes the spatially anisotropic triangular lattice (SATL), where two of the couplings are equal. Similar to the SATL, the SCATL can be tuned between vanishing and strong frustration.

This specific model may be relevant to understand the spin liquids occurring in related systems. E.g., in a previous work on the SATL and the square-lattice $J_1J_2J_3$ model (which has NN, next-NN, and next-next-NN interactions) \cite{Hauke2011}, we found that spiral long-range order (LRO) is never connected directly to collinear order. Instead the system seems to pass through a quantum disordered region. 
In this article, we want to examine if such a behavior remains true in the more general SCATL, to further corroborate if it is a general characteristic of two-dimensional frustrated spin systems. 

A related advantage of the chosen geometry is the possibility, given by the additional anisotropy, to approach the putative non-magnetic phases from different angles, possibly revealing crucial information not only about their location in parameter space, but also about their nature. Indeed, the results presented in this paper suggest that two gapped non-magentic regions, previously identified as two distinct phases in the SATL, might actually be continuously connected via the additional anisotropy of the SCATL.
Studying the persistence and characteristics of the putative quantum-disordered phases with respect to this additional anisotropy is the first main aim of the present paper. 

The second main aim is related to experimental findings in magnetic materials. While the SATL has found considerable attention in recent years, to our knowledge the ground-state phase diagram of the Heisenberg SCATL has never been thoroughly investigated. Recent first-principles calculations, however, show that some magnetic materials, such as the organic salts Me$_{4-n}$Et$_n$$Pn$[Pd(dmit)$_2$]$_2$ (abbreviated $Pn-n$) \cite{Scriven2011}, TMTTF \cite{Yoshimi2012}, or BaAg$_2$Cu[VO$_4$]$_2$ \cite{Tsirlin2011a}, which are well described by weakly-coupled 2D triangular lattices, can have considerable anisotropies between all three intra-plane couplings \footnote{Although in the last material the physics is dominated by a superposition of AFM and ferromagnetic 1D chains.}.
Typically, to locate the material within the well-studied SATL model, the two closer bond strengths are averaged. 
However, this places materials such as Sb-0 and As-2, which are experimentally found to be AFM ordered, into a region of the phase diagram, where according to many theoretical studies \cite{Weng2006,Yunoki2006,Heidarian2009,Hauke2011,Reuther2011} no LRO should exist.
The second aim of this paper, therefore, is to show that the additional anisotropy between the couplings could naturally explain this discrepancy, since it shifts the parameter values corresponding to these materials into an ordered phase. This also suggests that the non-magnetic state is quite sensitive to this additional anisotropy, which therefore has to be taken into account when interpreting experiments. 

In this work, we investigate the $S=1/2$ Heisenberg AFM SCATL, Eq.~\eqref{eq:HS}, within Takahashi's modified spin-wave theory (MSWT) \cite{Takahashi1989}, supplemented with the optimization of the ordering vector \cite{Hauke2010,Hauke2011}. 
Previously \cite{Hauke2010,Hauke2011}, we have shown that this improves significantly over conventional spin-wave theory (as well as over conventional MSWT), as it allows to account for the dramatic quantum corrections to the type of order appearing in frustrated quantum antiferromagnets. Further, the breakdown of the theory provides a strong signal that the true ground state might be quantum disordered; hence, this method serves to efficiently find candidate models for spin-liquid behavior. 
While the main focus of this article is on the MSWT results, we complement them with exact diagonalization (ED) of small clusters.
The SCATL with XY interactions, motivated by recent experiments with frustrated bosonic atoms in optical lattices, will be treated in a similar way in the following article~\cite{Hauke2012b}.

The rest of this paper is organized as follows. 
First, to understand which effects can be expected in the quantum SCATL, we discuss the phase diagram of its classical counterpart (Sec.~\ref{cha:phd_classical}) and briefly summarize known results from its well-studied limiting case, the SATL (Sec.~\ref{cha:KnownResults}). 
Sec.~\ref{cha:phd_quantum} contains the main results of our paper, namely the discussion of the quantum-mechanical ground-state phase diagram of the SCATL, including various observables from MSWT and ED as well as, for a possible comparison to experiment, the spin-wave dispersions at selected points of the phase diagram. 
We delegate the technical details of the MSWT to the Appendix.
Sec.~\ref{cha:conclusion}, finally, provides some conclusions.

\subsection{Classical phase diagram\label{cha:phd_classical}}

In this section, we discuss the classical phase diagram of the SCATL, which can serve as a guide to what ordered phases are to be expected, and which allows to appreciate the changes brought about by quantum fluctuations. 

To obtain the classical solution, we replace the Heisenberg spins in Eq.~\eqref{eq:HS} by classical rotors, which -- without loss of generality -- lie in the $xy$-plane \footnote{This allows us to neglect the $z$-component from now on.}. The ordering vector $\vect{Q}^{\mathrm{cl}}=\left(Q_{x}^{\mathrm{cl}},Q_{y}^{\mathrm{cl}}\right)$ is the $\vect{k}$-vector which minimizes the Fourier transform of the coupling strengths. It fixes the direction of each spin (up to a global phase) as 
$\vect{S}_{\vect{i}}=S\left(\cos(\vect{Q}^{\mathrm{cl}}\cdot\vect{r}_i),\sin(\vect{Q}^{\mathrm{cl}}\cdot\vect{r}_i)\right)$. 
We find
\eqa{
  Q_{x}^{\mathrm{cl}}& = &\left\{\begin{array}{l}
                              \pi \quad \mathrm{for} \quad -\frac{J}{2J'}-\frac{J'}{2J}+\frac{J J'}{2J''^2} \leq 1 \\
			      0   \quad \mathrm{for} \quad -\frac{J}{2J'}-\frac{J'}{2J}+\frac{J J'}{2J''^2} \geq 1 \\
			      \arccos\left( -\frac{J}{2J'}-\frac{J'}{2J}+\frac{J J'}{2J''^2} \right) \quad \mathrm{else}
                             \end{array}
			\right. \\
  Q_{y}^{\mathrm{cl}}& = &\left\{\begin{array}{l}
                              \pi \quad \mathrm{for} \quad -\frac{J}{2J''}-\frac{J''}{2J}+\frac{J J''}{2J'^2} \leq 1 \\
			      0   \quad \mathrm{for} \quad -\frac{J}{2J''}-\frac{J''}{2J}+\frac{J J''}{2J'^2} \geq 1 \\
			      \arccos\left( -\frac{J}{2J''}-\frac{J''}{2J}+\frac{J J''}{2J'^2} \right) \quad \mathrm{else}
                             \end{array}
			\right.
			      \nonumber
}

\begin{figure}
	\centering
	\includegraphics[width=0.49\textwidth]{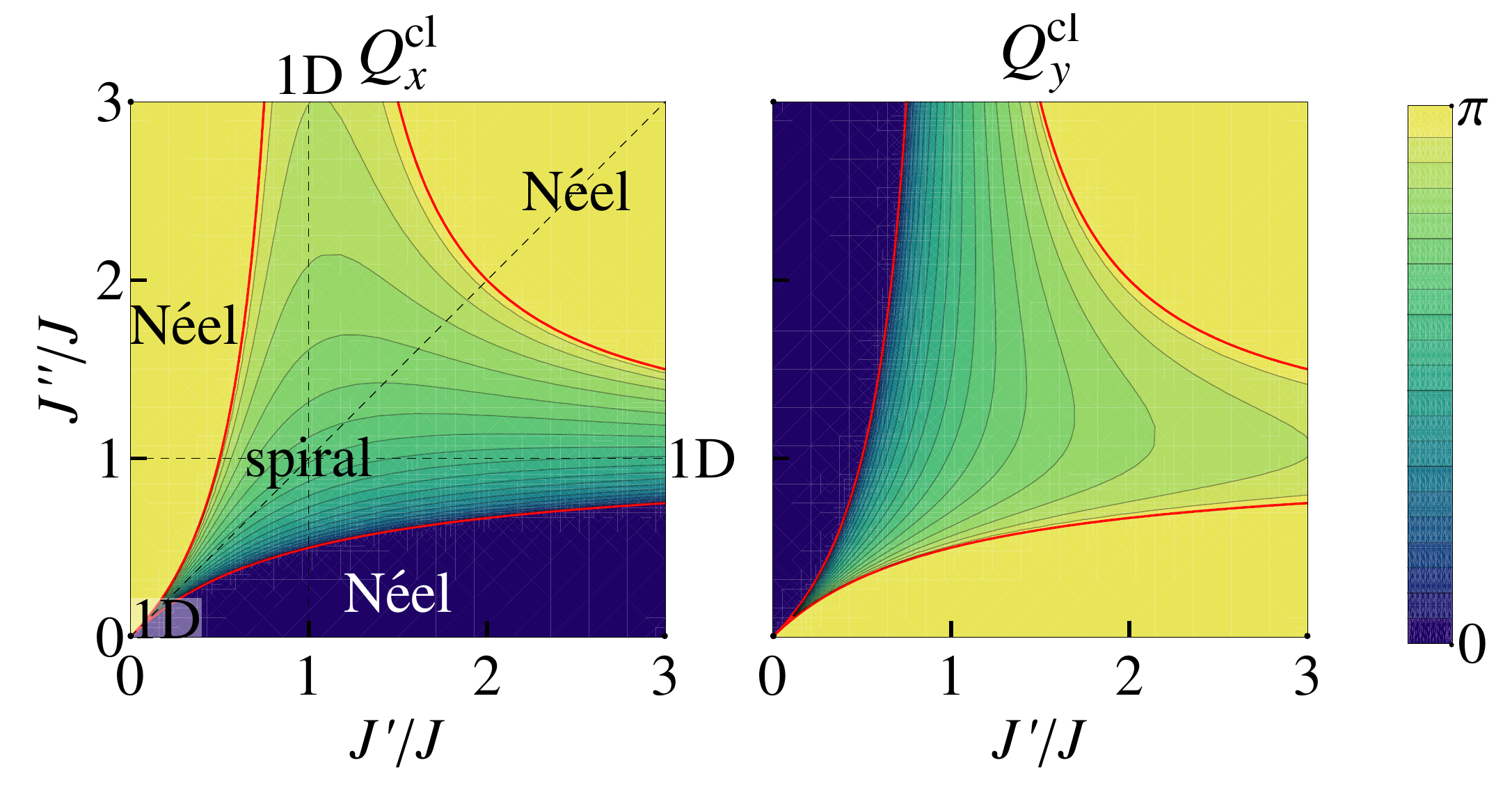}
	\caption{
	  {\bf Classical phase diagram of the SCATL: The ordering vector} evidences three N\'eel-ordered phases, an extended spiral-ordered phase, and limits where the system decouples into an ensemble of independent chains, as indicated by the labels in the left panel.
	  The thick red lines denote transitions between different kinds of order, and along the dashed black lines the system is in SATL limit.
	  \label{fig:phd_classical}
	}
\end{figure}

The classical phase diagram of the SCATL, plotted in Fig.~\ref{fig:phd_classical}, contains several N\'eel-ordered phases and an extended spiral-ordered phase. 
The N\'eel phases spread around the square-lattice limits $(J'/J,J''/J)=(1, 0)$ with $\vect{Q}^{\mathrm{cl}}=\left(0,\pi\right)$, $(J'/J,J''/J)=(0, 1)$ with $\vect{Q}^{\mathrm{cl}}=\left(\pi,0\right)$, and $J'/J,J''/J\gg1$ with $\vect{Q}^{\mathrm{cl}}=\left(\pi,\pi\right)$. 
The spiral phase, with continuously varying ordering vector, connects smoothly to the N\'eel phases, and occupies the extended region between them. In particular, it extends all the way to $J'/J=J''/J=0$ [and, symmetrically, to ($J'/J=1$, $J''/J\to\infty$) and ($J''/J=1$, $J'/J\to\infty$)], where the system decouples into an ensemble of 1D chains.

\subsection{Known results in limiting cases\label{cha:KnownResults}}

In this section, we discuss well-known limiting cases of the quantum SCATL, including results on the SATL. This helps us to assess which phases and quantum effects are to be expected in the phase diagram of the SCATL.

For $J'/J,J''/J\gg1$, $(J'/J,J''/J)=(1, 0)$, and $(J'/J,J''/J)=(0, 1)$, one recovers the square lattice limit. Here, N\'eel order persists also in the quantum case \cite{Richter2010}. Similarly, in the isotropic triangular lattice, $J'=J''=J$, spiral LRO survives quantum fluctuations \cite{Capriotti1999a}. 
The limits ($J'=J''=0$), ($J'\to\infty$ with $J''=\mathrm{const}$), and ($J''\to\infty$ with $J'=\mathrm{const}$) correspond to ensembles of decoupled, critical Heisenberg chains, with algebraic correlations along individual chains but no correlations between them.

\begin{figure}
	\centering
	\includegraphics[width=0.49\textwidth]{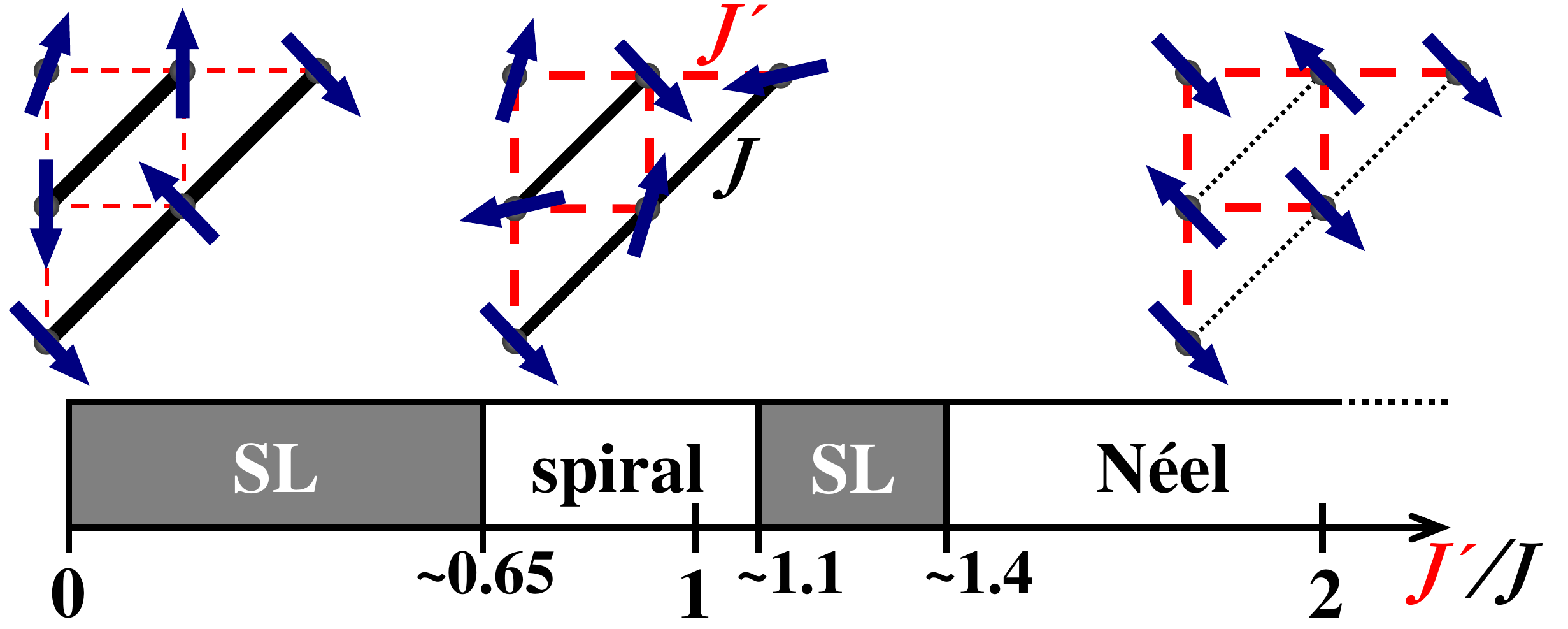}
	\caption{
		{\bf The MSWT quantum phase diagram of the SATL} (from Ref.~\cite{Hauke2011}) contains N{\'e}el order (which is considerably more stable than in the classical model), spiral order (which is destabilized by quantum fluctuations), and two putative spin-liquid (SL) phases. These are found through the breakdown of the theory and a disappearing  spin stiffness, which indicates a gapped disordered phase. In the purely 1D limit, MSWT recovers convergence and produces the 1D critical state. 
		We include sketches of classical states, where blue arrows indicate the directions of the classical rotors, namely, the 1D state at $J'/J=0$, the spiral state at $J'/J=1$, and the 2D-N\'{e}el state at $\alpha\geq 2$, . 
	  \label{fig:phd_SATL}
	}
\end{figure}

For $J'=J''\equiv\alpha J$ (or, equivalently, $J'=J$ or $J''=J$), one recovers the SATL, which is realized in a variety of $S=1/2$ compounds, e.g., $\mathrm{Cs}_2\mathrm{CuCl}_4$ \cite{Coldea2001} and $\kappa$-(BEDT-TTF)$_2$Cu$_2$(CN)$_3$ \cite{Shimizu2003, Yamashita2008}. The model may display spin-liquid phases, although their extent and nature is still under intensive debate \cite{Weihong1999,Weng2006,Yunoki2006,Kohno2007,Fjaerestad2007,Starykh2007,Heidarian2009,Hauke2011,Weichselbaum2011,Reuther2011,Ghamari2011}.
In the rest of this section, we review the main features of the SATL phase diagram as found in the literature, proceeding from large to small $\alpha\equiv J'/J=J''/J$ (for comparison, Fig.~\ref{fig:phd_SATL} reproduces the MSWT phase diagram from Ref.~\cite{Hauke2011}). 

It is commonly accepted that order-by-disorder effects due to quantum fluctuations stabilize the N{\'e}el phase considerably over the classical model, moving the point where N{\'e}el order disappears downwards from the classical value $\alpha=2$ to values between $\alpha\approx1.1$ and $1.67$, depending on the method used \cite{Weihong1999,Manuel1999,Hauke2011,Weichselbaum2011,Reuther2011,Doretto2012}.
Further, several methods predict that quantum fluctuations spread the transition point between the N{\'e}el and the spiral phase into a quantum-disordered phase \cite{Weihong1999,Weng2006,Hauke2011}. In the following, we term this predicted disordered region ``large-$\alpha$ quantum-disordered region'' (large-$\alpha$ QDR). 
One of the main aims of this article is to study if it is a general feature of frustrated quantum antiferromagnets that a quantum-disordered phase intervenes in transitions between commensurate and incommensurate order. 

Similar behavior has been found in a variety of quantum spin models, including $J_1J_2J_3$-models on the square lattice \cite{
Chandra1988,Locher1990,Ferrer1993,Zhong1993,Leung1996,Capriotti2004a,Capriotti2004b,Mambrini2006,Shannon2006,
Murg2009,Richter2010,Hauke2011,Reuther2011b},  
and frustrated honeycomb models with Heisenberg
\cite{Farnell2011,Albuquerque2011,Li2012}
or XY interactions
\cite{Varney2011}. 
In fact, since quantum phase transitions are driven by quantum fluctuations, one might expect that -- if anywhere -- a complete restructuring of the ground state in favour of a quantum mechanical configuration may occur preferably close to a quantum critical point. 
It, hence, seems plausible that at such points quantum fluctuations are most effective in disrupting classical order. 
Indeed, a similar effect occurs in classical statistical physics. Assume that there is a transition between a commensurate and an incommensurate phase which both show LRO. As a first notable thing, close to this transition the thermal phase transition to a disordered state will typically happen at lower temperature than far away from it. Beyond the thermal phase transition, the disordered phase will show short-range order of the type corresponding to the adjacent long-range ordered phase. The transition between the two different kinds of short-range order is called a \emph{disorder point} \cite{Stephenson1969,Stephenson1970,Stephenson1970a,Stephenson1970b}. 
Interestingly, the correlation length associated with the two kinds of short-range order has a minimum just at this point. 
Hence -- similar to what is found in the quantum models -- thermal fluctuations tend to suppress order most effectively at a commensurate--incommensurate transition. 

At the low-$\alpha$ side of the spiral phase, previous works predict a disordered phase, which could appear for as large $\alpha$ as $\approx 0.8-0.9$ \cite{Weng2006,Yunoki2006,Heidarian2009,Reuther2011}.  Our previous MSWT results suggest $\alpha\approx0.65$ \cite{Hauke2011}.
In the following, we term this predicted disordered region ``small-$\alpha$ QDP.''
It may be associated to a spread of the gapless spin liquid of the isolated chains ($J' = 0$) to finite coupling \cite{Yunoki2006,Heidarian2009,Reuther2011}, possibly followed by a gapped spin liquid \cite{Yunoki2006,Heidarian2009,Hauke2011}.
This double nature of the disordered region is still under debate, since some works only find a gapless spin liquid \cite{Reuther2011}. Consistent between these methods is the prediction that quantum fluctuations disrupt ordering tendencies between the chains even for relatively large inter-chain couplings. 
But consent about the physics in this region seems far from reached. For example, recent DMRG studies entirely question the existence of the small-$\alpha$ spin liquid(s) \cite{Weichselbaum2011}. And a recent renormalization-group analysis \cite{Ghamari2011} found collinear AFM long-range order in the region $\alpha\leq0.3$ (see also \cite{Starykh2007}), and, above that value, spiral order. The weakness of the spiral order leaves, however, the possibility that the true quantum ground state hosts a disordered phase in the parameter range $0.3-0.5$.

\section{Quantum-mechanical phase diagram\label{cha:phd_quantum}}

From the discussion of the classical phase diagram and the limiting cases, we have the necessary background to tackle the quantum-mechanical ground-state phase diagram of the Heisenberg SCATL. 
To compute it, we use the MSWT supplemented with ordering-vector optimization, working directly in the thermodynamic limit.
Since this method is described in detail in our previous articles \cite{Hauke2010,Hauke2011}, we delegate the technical aspects to the Appendix, and only summarize here the main idea. The starting point is a classical state, which one dresses with quantum fluctuations in a second-order spin-wave expansion. This yields a bosonic Hamiltonian, the ground state of which is found self-consistently by minimizing its mean-field free energy. For this, quartic terms, i.e., interactions between spin waves, are decoupled via Wick's theorem. 
Additionally, we employ Takahashi's modification of vanishing magnetization. This constricts the average number of spin-wave excitations to a physical value, in contrast to conventional spin-wave theory, where the spin-wave excitations grow completely unchecked. 
This modification has proven a crucial improvement to describe low-dimensional systems with weak order tendencies. 

Typically, one uses the classical ground state as the reference state. However, in many models quantum fluctuations considerably shift the type of predominant order. Therefore, we find the ordering vector giving the best classical reference state by including it in the self-consistent optimization. This has proven crucial to capture, e.g., the stabilization of the N\'eel phase by quantum fluctuations. 
As has been proposed in Refs.~\cite{Hauke2010,Hauke2011}, the breakdown of the theory strongly suggests that at mean-field level \emph{no} semi-classical reference state yields a good description of the quantum ground state. This is then interpreted as an indication of non-magnetic behavior in the true ground state. This will be an important aspect for the interpretation of the quantum phase diagram. 

We compare these MSWT results to exact diagonalization (ED) of a 15-site lattice, as depicted in Fig.~\ref{fig:geometry}. 
The geometry is chosen for its symmetry between $J$, $J'$, $J''$ bonds. 
It is important to leave the boundaries open to allow for incommensurate ordering vectors.

\subsection{MSWT and ED results -- ordering vector and order parameter \label{cha:quantumPhaseDiagramResultsEDandMSWT}}

In this section, we give a first overview over the phase diagram, as obtained from the ordering vector $\vect{Q}$ and the order parameter $M$, followed in the next two sections by more detailed analyses. 
In MSWT, ordering vector and order parameter are direct results of the optimization [see Appendix, Eqs.~\eqref{eq:orderParameterMSWT} and~\eqref{Qs}]. In ED, they can be extracted from the static structure factor, 
\eq{
S(\vect{k})=\frac{1}{N^2}\sum_{i,j} \ue^{i \vect{k}\cdot(\vect{r}_i-\vect{r}_j)} \braket{\vect{S}_i\cdot\vect{S}_j}\,.
}
Its peak lies at the ordering vector $\vect{Q}^{\mathrm{ED}}$, and the square root of its height, $\sqrt{S(\vect{Q}^{\mathrm{ED}})}\equiv M^{\mathrm{ED}}$, approaches in the thermodynamic limit the order parameter $M$. 

\begin{figure}
	\centering
	\hspace*{-0.5cm}\includegraphics[width=0.52\textwidth]{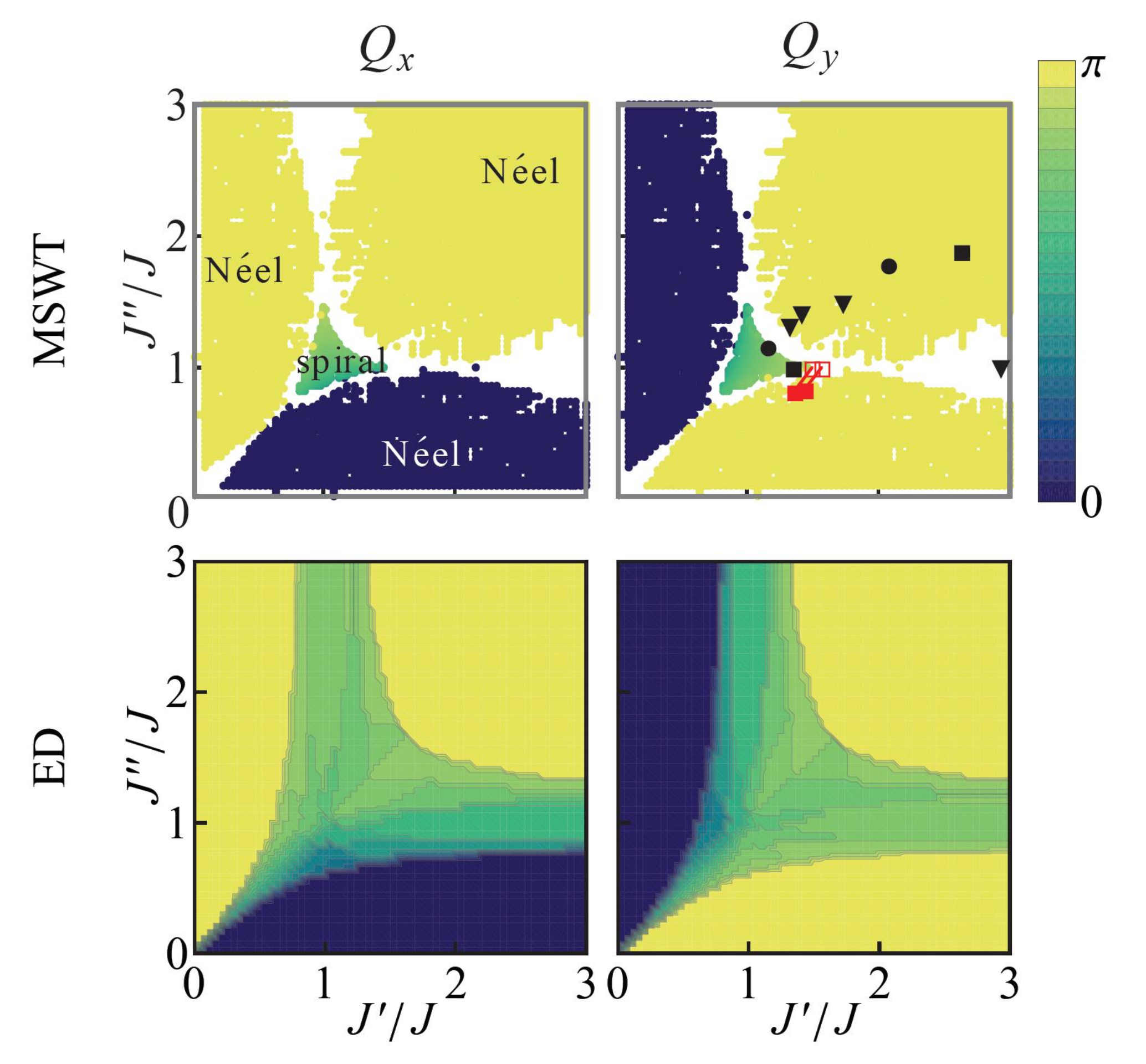}
	\caption{
	  {\bf Quantum-mechanical phase diagram of the SCATL, ordering vector.} 
	  {\bf Upper row: MSWT data.} Quantum fluctuations stabilize the N\'eel phase. Around $J'\approx J''\approx J$, a part of the classical spiral phase survives quantum fluctuations (labels in the upper left panel). 
	  The solid symbols (upper right panel) denote some experimental materials (see Table~\ref{tab:materials}). (For clarity, we show only symbols in the lower right part of the figure, excluding points symmetric under exchange of $J$, $J'$, and $J''$.)
	  $\filledmedtriangleup$: magnetically disordered, 
	  {\Large{\textbullet}}: charge ordered, 
	  $\filledmedsquare$: AFM LRO. 
	  Note especially the two {\color{red}$\filledmedsquare$} at $(J'/J,J''/J)=(1.44, 0.84)$ and $(J'/J,J''/J)=(1.36, 0.82)$ marking the materials As-2 and Sb-0, which lie well inside a N\'eel ordered phase. Neglecting the asymmetry between the couplings would put As-2 into the supposedly disordered region and Sb-0 just at its boundary ({\color{red}$\medsquare$}). 
	  {\bf Lower row: ED data} for $N=15$ sites. Already for this small system, it can be appreciated that (compared to the classical case) the N\'eel phase grows at the expense of spiral order.
	  \label{fig:phd_quantum_Q}
	}
\end{figure}
\begin{figure}
	\centering
	\hspace*{-1cm}\includegraphics[width=0.55\textwidth]{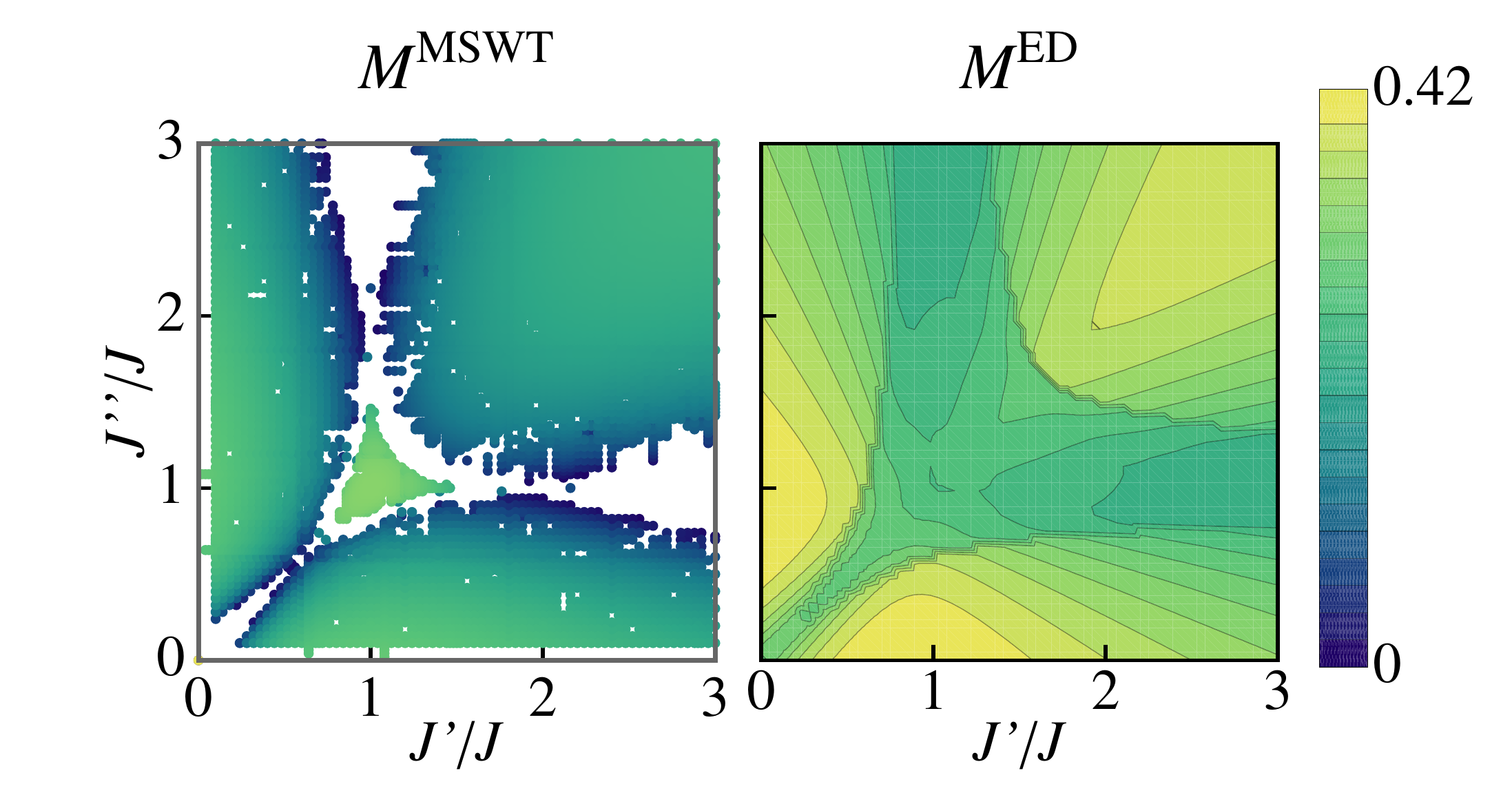}
	\caption{
	  {\bf Quantum-mechanical phase diagram, order parameter.} ED results qualitatively confirm MSWT. In particular, the order parameter for both methods decreases rapidly upon approaching the MSWT breakdown regions. 
	  \label{fig:phd_quantum_M}
	}
\end{figure}
As seen in the MSWT and ED ordering vectors, presented in Fig.~\ref{fig:phd_quantum_Q}, quantum fluctuations stabilize the N\'eel phases compared to the classical case, as already observed in the SATL (Sec.~\ref{cha:KnownResults}).
In the central region around $J'\approx J''\approx J$, a broad range of incommensurate ordering vectors indicates spiral order. 
The finite MSWT order parameter (Fig.~\ref{fig:phd_quantum_M}, left panel) shows that in these phases indeed LRO survives quantum fluctuations.
(Note that the self-consistent MSWT calculations become relatively unstable for small order parameters, which results in ragged phase boundaries.)

In the N\'eel phases, the ED order parameter (Fig.~\ref{fig:phd_quantum_M}, right panel) is maximal, giving support to the assumption that here LRO persists. However, it is much smaller in the spiral phase than the MSWT value, a discrepancy already found in the SATL~\cite{Hauke2011}. (This could be due to third-order corrections to the spin-wave expansion which our approach neglects, and which can become important in spiral configurations  \cite{Chernyshev2009}.)

Between the ordered phases, we find a broad region where MSWT breaks down, indicating as usual \cite{Hauke2010,Hauke2011} that these regions do not allow a description in terms of an ordered, semi-classical state. 
Therefore, it appears that it is a quite universal feature of frustrated quantum antiferromagnets that spiral- and collinearly-ordered phases are always separated by quantum disordered phases. This is the first main result of this paper. 

The strong decrease of the MSWT and ED order parameters (Fig.~\ref{fig:phd_quantum_M}) upon approaching this region gives support to this interpretation (which we will further corroborate in the next two sections). 
Note also that both the ED and MSWT order parameter seem to disappear more smoothly when approaching the putative 1D-like QDR (consider, e.g., in the range $2\lesssim J'/J\lesssim 3$, $J''/J\to 1^-$). Upon approaching the putative large-$\alpha$ QDR dividing spiral from N\'eel LRO, on the other hand, for ED, the order parameter decreases sharply (consider, e.g., the line $J'/J=1$, $J''/J\to 1^-$).  Here, for MSWT, the breakdown occurs abruptly at finite order parameters. 
This could point at a difference in the type of phase transition upon approaching the large-$\alpha$ QDR and the non-magnetic phase at the decoupled-chains limit.

The second main result of our paper concerns experimental measurements of ground-state behavior of some materials, taken from Ref.~\cite{Scriven2011} (see also the reviews~\cite{Kanoda2011,Powell2011}), as well as from Refs.~\cite{Coldea2002} (Cs$_2$CuCl$_4$) and~\cite{Ono2005} (Cs$_2$CuBr$_4$).
For reference, they are presented in Table~\ref{tab:materials}, and included as solid symbols in the upper right panel of Fig.~\ref{fig:phd_quantum_Q}.
\begin{table}
 \begin{tabular}{ l c r }
  \toprule
  		material & $(J'/J, J''/J)$ & state  \\
  \midrule
  		N-3 & (10, 9.1) & AFM \\
  		P-2 & (2.63, 1.89) & AFM \\
  		Sb-2 & (2.08, 1.79) & CO \\
  		Sb-1 & (1.72, 1.49) & SL \\
  		$\kappa$-CN & (1.41, 1.41) & SL \\
  		P-1 & (1.32, 1.32) & VBS \\
  		Cs & (1.16, 1.16) & CO \\
  		Sb-0 & (0.74, 0.60) & AFM \\
  		Cs$_2$CuBr$_4$ & (0.74, 0.74) & AFM  \\
  		As-2 & (0.69, 0.58) & AFM \\  		
  		Cs$_2$CuCl$_4$ & (0.34, 0.34) & RVB	\\
  \bottomrule
\end{tabular}
\caption{
{\bf Some relevant materials} for which the ground state has been measured in experiment, together with the coupling strengths, and the state they are found to be in (from Refs.~\cite{Scriven2011,Coldea2002,Ono2005} and references therein). AFM: antiferromagnetic LRO, CO: charge ordered, SL: spin liquid, VBS: valence-bond solid, RVB: resonating valence-bond state. 
\label{tab:materials}}
\end{table}
We mark magnetically disordered materials (spin liquids, resonating valence-bond states, or valence-bond solids) with triangles, charge-ordered materials with bullets, and AFM-ordered materials with squares. 
Including the full anisotropy of the triangular lattice, all AFM ordered materials lie within the ordered phases from MSWT \footnote{On the other hand, some magnetically disordered materials lie in ordered regions of the MSWT phase diagram. It is known, however, that MSWT overestimates ordered phases.}.
In particular, the AFM ordered materials As-2 and Sb-0 [at $(J'/J,J''/J)=(1.44, 0.84)$ and $(J'/J,J''/J)=(1.36, 0.82)$] lie well inside a N\'eel ordered phase. If one neglects the anisotropy between $J'$ and $J''$, taking as usual the mean of both couplings, they would lie at the position of the empty squares at $(J'/J,J''/J)=(1.57, 1)$ [equivalent to $(J'/J,J''/J)=(0.64,0.64)$] and $(J'/J,J''/J)=(1.50, 1)$ [equivalent to $(J'/J,J''/J)=(0.67,0.67)$] -- inside a phase where many methods \cite{Yunoki2006,Weng2006,Heidarian2009,Reuther2011} predict disorder; specifically, within MSWT, symmetrizing the couplings puts Sb-0 just at the border to the breakdown region (which should be a lower limit for a disordered phase in the true ground state) and As-2 within it. 
The appearance of AFM N\'eel LRO in these experiments may find, therefore, a simple explanation in the full anisotropy of the SCATL. This second main result of our paper shows how crucial the full anisotropy is for the interpretation of experimental data. 
The rest of this article is devoted to fleshing these main findings out.

\subsection{Supporting observables from MSWT -- spin stiffness and spin-wave velocities \label{cha:spinStiffnessAndSpinWaveVelocities}}

\begin{figure}
	\centering
	\hspace*{-3cm}\includegraphics[width=0.35\textwidth]{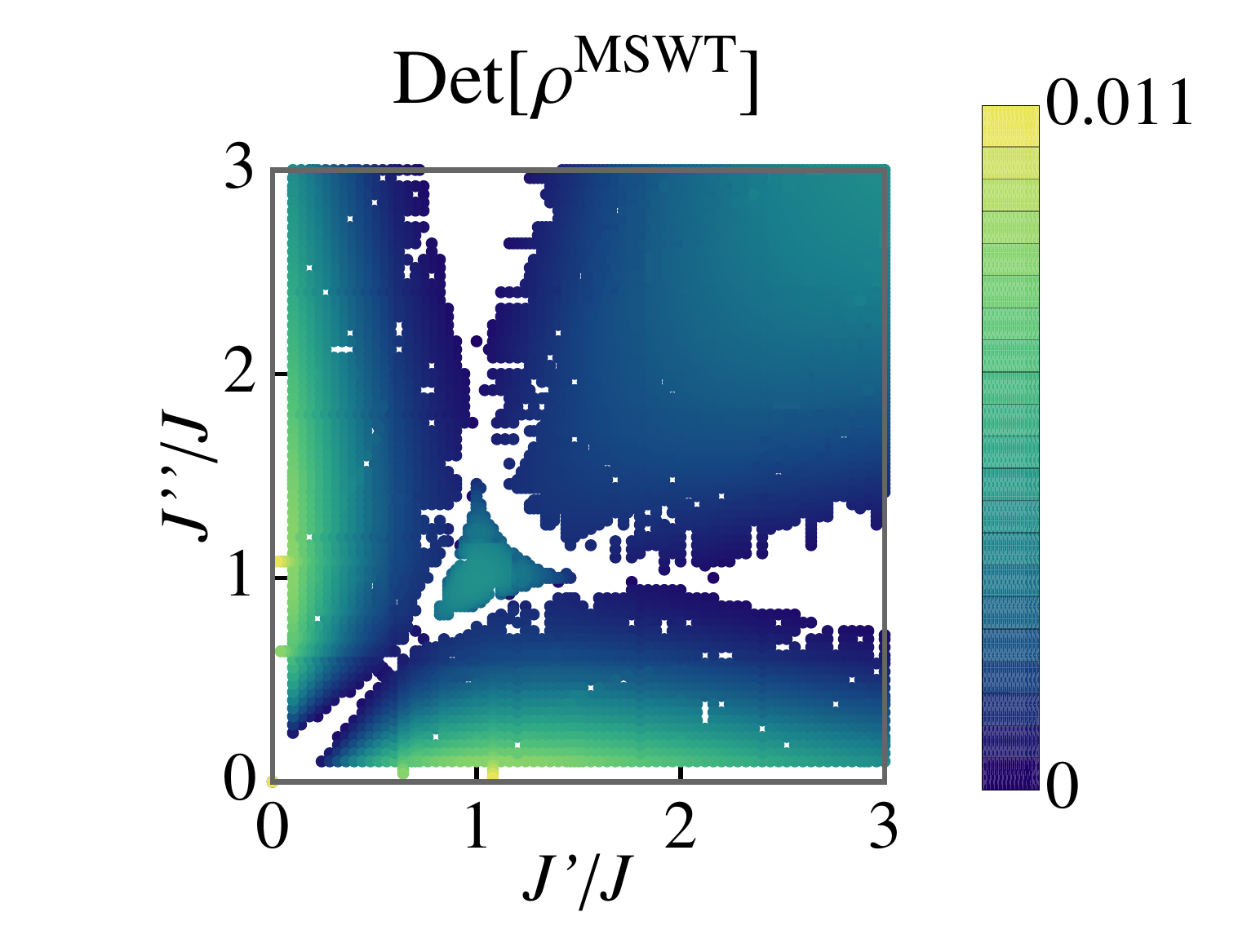}\\
	\includegraphics[width=0.49\textwidth]{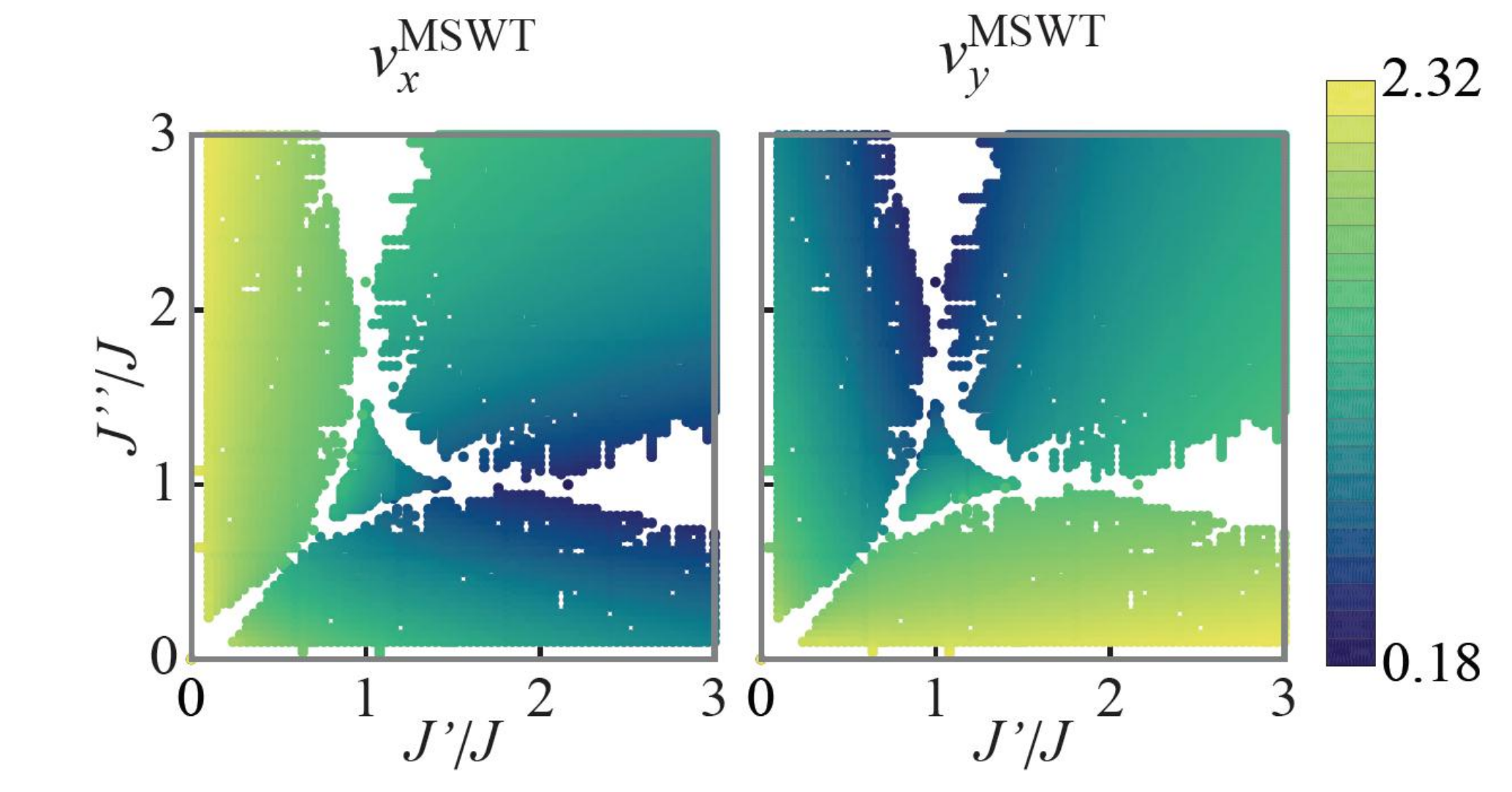}
	\caption{
	  {\bf Upper panel: The partial spin stiffness} decreases upon approaching the MSWT breakdown region, suggesting the disruption of magnetic LRO.
	  {\bf Lower panels: The spin-wave velocities} perpendicular to the dominating coupling strength soften in the 1D limits. Differences in the spin-wave velocities might allow to measure the anisotropy of the SCATL.
	  All quantities are normalized to the coupling strengths $1+J'+J''$. 
	  \label{fig:phd_MSWT_spinvelocityAndStiffness}
	}
\end{figure}
In Refs.~\cite{Hauke2010,Hauke2011}, the spin-stiffness tensor, which characterizes the stiffness of the magnetic order under change of the ordering vector, has proven a valuable consistency check of our MSWT calculations. 
Its components are 
\eq{
  \label{eq:spinstiffness}
  \rho_{\alpha,\beta}=\frac{d^2 \mathcal{F}}{d Q_\alpha d Q_\beta}\,,
}
where $\mathcal{F}$ is the free energy. Even if the order parameter is finite, a small spin stiffness suggests that further quantum fluctuations than taken into account within MSWT could disrupt the remaining order
\footnote{Also, if the spin stiffness is small when approaching the MSWT breakdown region, we are led to assume that the breakdown is not caused by numerical problems, but that it is a physical effect, i.e., due to the lack of a description in terms of a semi-classically ordered reference state.}.

Since for our purposes an upper bound for the spin stiffness is sufficient, we take the partial derivative in Eq.~\eqref{eq:spinstiffness}. The exact spin stiffness can be computed via the total derivative. To this, within the self-consistent MSWT calculations, one first has to find the optimal ordering vector. Then, one reruns the self-consistent MSWT equations for several fixed, slightly non-optimal ordering vectors, yielding slightly larger energies. The spin stiffness can be derived by fitting a quadratic form to the resulting energy landscape. In the self-consistent iteration, the mean fields characterizing the MSWT state can adjust to a changed ordering vector. This effect is not taken into account in the partial derivative, which hence provides an upper bound to the total spin stiffness. We find that it suffices to extract the location of disordered phases, but it may yield wrong results about their nature. 
In particular, we found in the SATL \cite{Hauke2011} that, upon approaching the putative small-$\alpha$ QDR, not only the total inter-chain, but also the total intra-chain spin stiffness decreases strongly. Since such a behavior is not consistent with algebraic correlations along the chains, this can be interpreted as an indication of a gapped quantum-disordered state. The partial spin stiffness computed in Ref.~\cite{Hauke2011}, on the other hand, only vanishes in the inter-chain direction. Hence, it may not be able to distinguish gapped from gapless spin liquids. However, it still seems to adequately capture the location of disordered regions.

In Fig.~\ref{fig:phd_MSWT_spinvelocityAndStiffness}, upper panel, we show the determinant of the spin-stiffness tensor, $\det(\rho)$, normalized to the coupling strengths $1+J'+J''$.
As we should expect \cite{Chubukov1994}, $\det(\rho)$ decreases upon approaching the phase transitions, especially from the N\'eel-ordered side.
At large $J'$ ($J''$), this decrease is due to a softening of the stiffness in $x$ ($y$) direction, and at small $(J'/J,J''/J)$ in the direction perpendicular to $\vect{\tau}_1$ (as has also been found in Ref.~\cite{Hauke2011}).

Another indicator for approaching disordered phases is given by the spin-wave velocities $v_{x,y}$, which can be connected to the spin stiffness via the susceptibility~\cite{Halperin1969}.
Since the spin-wave velocities are defined as the leading order of an expansion of the spin-wave dispersion relation, Eq.~\eqref{disp}, around small $\left|\vect{k}\right|$, i.e., 
\begin{subequations}
\eqa{
    v_x&=&\left.\lim_{k_x\to 0} \omega_{\vect{k}}/k_x\right|_{k_y=0} \,, \\
    v_y&=&\left.\lim_{k_y\to 0} \omega_{\vect{k}}/k_y\right|_{k_x=0} \,,
}
\end{subequations}
they can be measured directly from the spin-wave dispersion, allowing an experimental check of our findings. 

As seen in Fig.~\ref{fig:phd_MSWT_spinvelocityAndStiffness}, lower panels, close to the 1D breakdown region, they, too, soften in the direction perpendicular to the dominating coupling. 
On the other hand, when approaching the putative large-$\alpha$ QDR dividing the spiral from the N\'eel phase, both spin-wave velocities remain finite. This is another (besides the different behavior of the order parameter) indication that the large-$\alpha$ QDR could be qualitatively different from the non-magnetic phase found in the limit of decoupled chains.

\subsection{Supporting observables from ED -- energy derivative, gap, and chiral correlations}

\begin{figure}
	\centering
	\hspace*{-1cm}\includegraphics[width=0.55\textwidth]{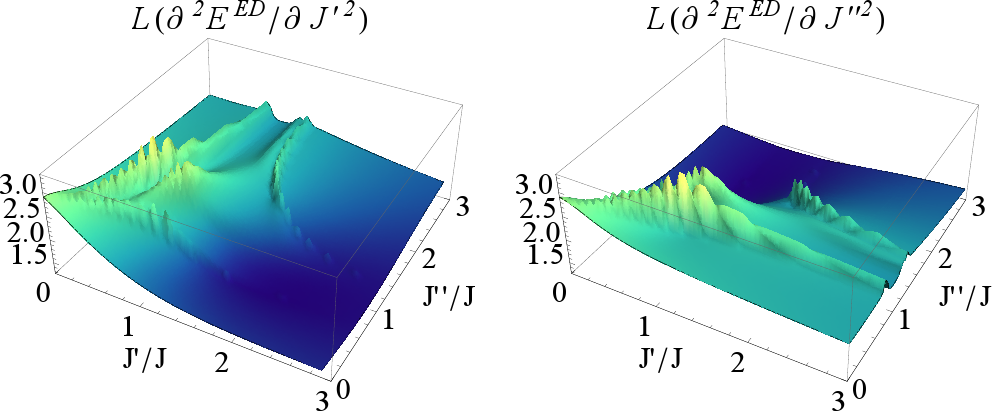}
	\caption{
	  {\bf Second derivative of ED ground-state energy} per spin for $N=15$. For clarity, we plot the logarithm after a shift to values larger one, $L(\partial^2 E^{\mathrm{ED}} / \partial {J^\gamma}^{2})$, where $L(x)=\log(1+\max(x)-x)$, and $J^{\gamma}=J'$ or $J''$. Strong peaks mark the transitions from the N\'eel phases. An additional peak around $(J',J'')=(1,1)$ might be an indication of an additional phase, separating the N\'eel phases from the spiral one. 
	  \label{fig:phd_ED_energy2ndDerivative}
	}
\end{figure}

The ED observables investigated in Sec.~\ref{cha:quantumPhaseDiagramResultsEDandMSWT} allowed to interpret the predominant ordering behavior, but did not yield clear evidence if within ED really quantum phase transitions exist, and if yes, where. 
The second derivatives of the ED ground-state energy per spin, plotted in Fig.~\ref{fig:phd_ED_energy2ndDerivative}, can provide such an indicator. In the thermodynamic limit, it diverges at a quantum phase transition. 

Indeed, there are clear peaks at lines similar to where in MSWT the N\'eel order breaks down. Also, a peak appears around $(J',J'')=(1,1)$. This might be a precursor of a quantum phase transition away from the spiral state, and to an intermediate phase possibly to the QDR which is supposed to exist in this system. 

\begin{figure}
	\centering
	\includegraphics[width=\columnwidth]{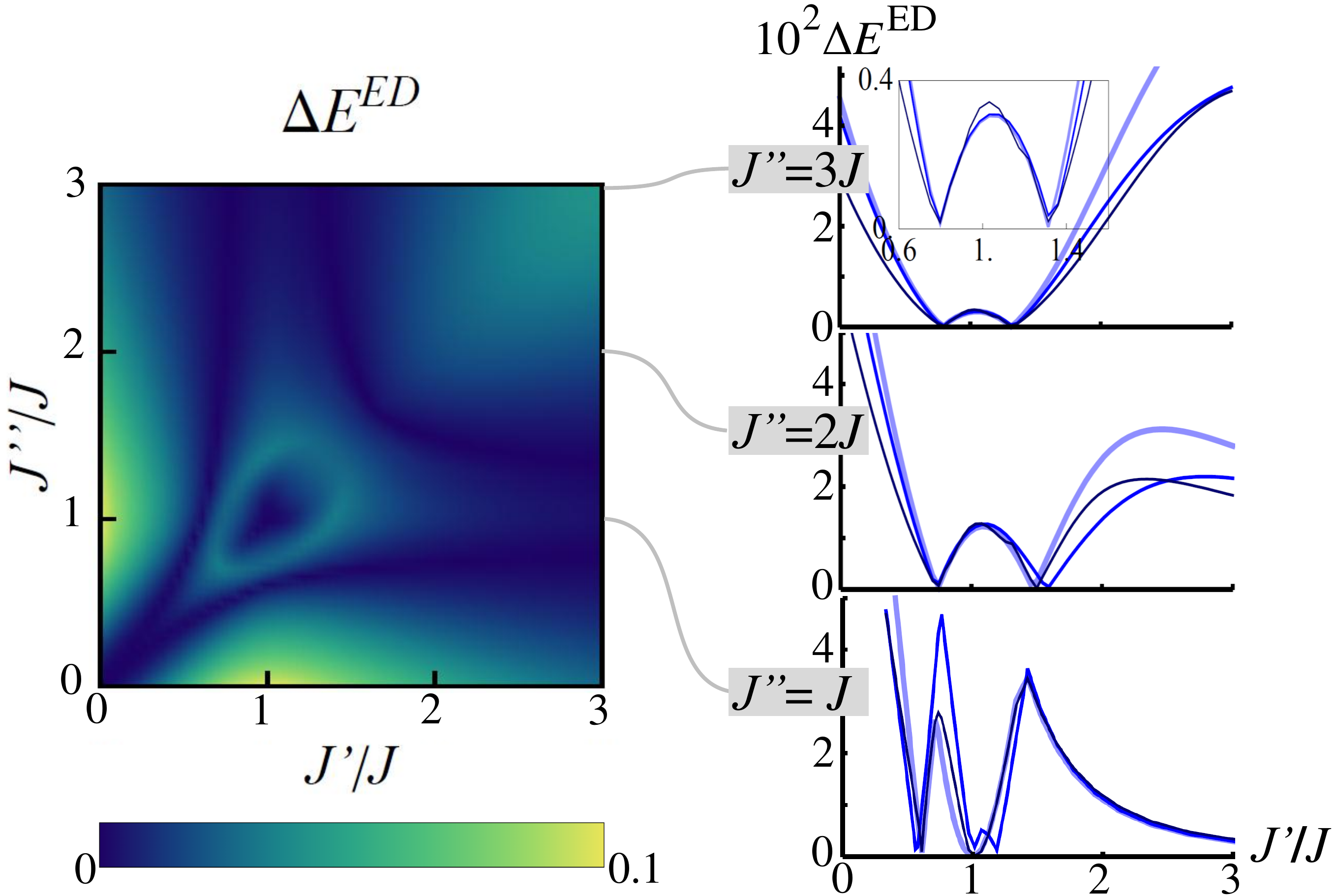}
	\caption{
	  {\bf Left panel: The ED gap per spin} lends support to the MSWT phase diagram: A finite gap separates in the N\'eel phases spin-wave excitations from the ground state. In the spiral phase, the ground state is doubly degenerate due to the ambiguity in choice of chirality. The finite gap surrounding the degenerate region could be a precursor of a gapped, disordered phase. At the quantum phase transitions to the N\'eel phases, the gap closes again. 
	{\bf Right panels: cuts} at fixed $J''/J=1,2,3$ for triangular systems with increasing $N$ (from light to dark and thick to thin: 6,10,15). There is little size dependence in the central gapped phase ($J''/J=2,3$ with $J'\approx J$, as well as $J''=J$ and $J'/J\gtrsim 1.5$). Also, for $J''=3J$ the transition points do not show any appreciable size dependence, while for $J''=2J$ the one around $J'/J=1.5$ does. 
	  \label{fig:phd_ED_gap}
	}
\end{figure}

We get further support for this phase diagram from the ED energy gap between ground and first excited state, Fig.~\ref{fig:phd_ED_gap}. 
In the well-known limiting cases of the SCATL, it behaves as expected: 
There is no singlet gap close to the decoupled-chains limits, since the system is then in a critical phase. 
In the N\'eel ordered phases, there is a large gap which separates the ground state from closely-spaced excitations, which in larger lattices become the spin waves, collapsing slowly towards the ground state \cite{Lhuillier2005}. 
This is consistent with the considerable size dependence found for these parameter regions, as can be seen in the right panels of Fig.~\ref{fig:phd_ED_gap}, where we plot cuts of the gap per spin, $\Delta E^{\mathrm{ED}}$, at fixed $J''/J=1,2,3$ for triangular systems similar to the one in Fig.~\ref{fig:geometry} with $N=6,10,15$.

On the contrary, there is no gap in the spiral phase, 
because there are two degenerate ground states with opposite chirality 
\footnote{In the spiral phase, there is a gap, similar to the spin-wave gap of the N\'eel phases, between the \emph{second} and the \emph{third} energy level.}. 
We find that this vanishing of the gap depends strongly on the system geometry, but it occurs consistently for all triangular systems considered.

Interestingly, the gapless spiral phase is surrounded by a region where the gap attains considerable values. The very small dependence on system size for this parameter region indicates that this is stable towards the thermodynamic limit.
A finite gap is not consistent with a spiral-ordered phase. On the other hand, the predominant order in this region is at incommensurate wave-vectors. Hence, the finite gap is clearly not  due to square-lattice N\'eel physics. 
Optimistically, these findings could therefore be interpreted as the precursors of a gapped QDR. 
This gapped region completely encircles the spiral phase, suggesting that the low- and large-$\alpha$ gapped QDRs found in the SATL could actually be continuously connected via the additional anisotropy of the SCATL. 

Upon approaching the N\'eel phases, the gap closes, indicating a quantum phase transition. 

Going back to Fig.~\ref{fig:phd_quantum_M}, when comparing MSWT and ED the lateral extent of the putative non-magnetic phases is different. Scanning along $J'=J$, within MSWT it is smallest around $J''=2J$, while for ED it decreases monotonously with increasing $J''$. 
As the right panels in Fig.~\ref{fig:phd_ED_gap} indicate, this discrepancy could be due to finite-size effects. Indeed, we find in ED that for $J''/J=2$ the transition point at $J'/J=1.5$ shows an appreciable size dependence, while the transition points for $J''/J=3$ do not. 
Therefore, the lateral extent of the putative QDRs at around $J'/J=2$ could decrease with $N$, making the MSWT and ED pictures consistent. 

From the gap, it seems that there is support for an extended gapped phase separating spiral and N\'eel LRO. Still, it would be desirable to exclude for this region spiral LRO in the thermodynamic limit. 
For this, we now study where chiral correlations persist.
The vector chirality is defined as 
\eq{
  \kappa_{i,j,k}=\frac{2}{3\sqrt{3}}\left(\vect{S}_i\times\vect{S}_j+\vect{S}_j\times\vect{S}_k+\vect{S}_k\times\vect{S}_i\right)_z\,,
}
where the sites $\left\{i,j,k\right\}$ are located counter-clockwise on a triangle. For the small systems used in our ED, we generalize the chiral correlations \cite{Richter1991} to
\eq{
  \Psi_{-}=\frac{4}{N_{\Delta}}\braket{\sum_c s_c \kappa_c \sum_a s_a \kappa_a}\,.
}
Here, the sum $a$ runs over all triangles, while $c$ runs only over the central ones to reduce boundary effects. The factors $s_{a,c}$ weight $\kappa_{a,c}$ with a $+$ ($-$) sign if the triangle points upwards (downwards). 
The prefactor, where $N_{\Delta}$ is the number of summands, is chosen such that the chiral correlation has the same theoretical maximum of $\frac{9}{4}$ as the usual definition for large lattices \cite{Richter1991}.

\begin{figure}
	\centering
	\includegraphics[width=0.49\textwidth]{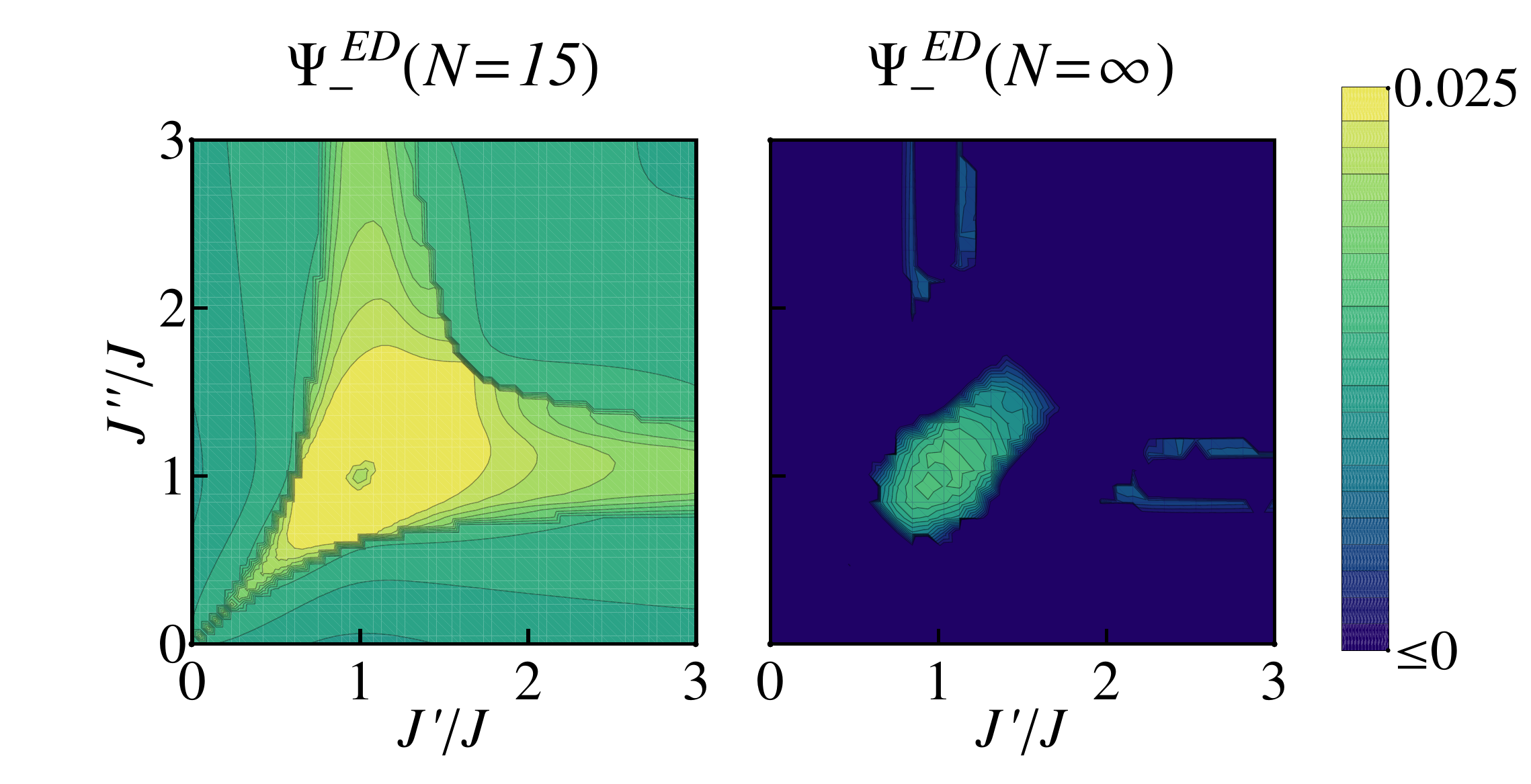}\\
	\hspace*{1.75cm}\includegraphics[width=0.4\textwidth]{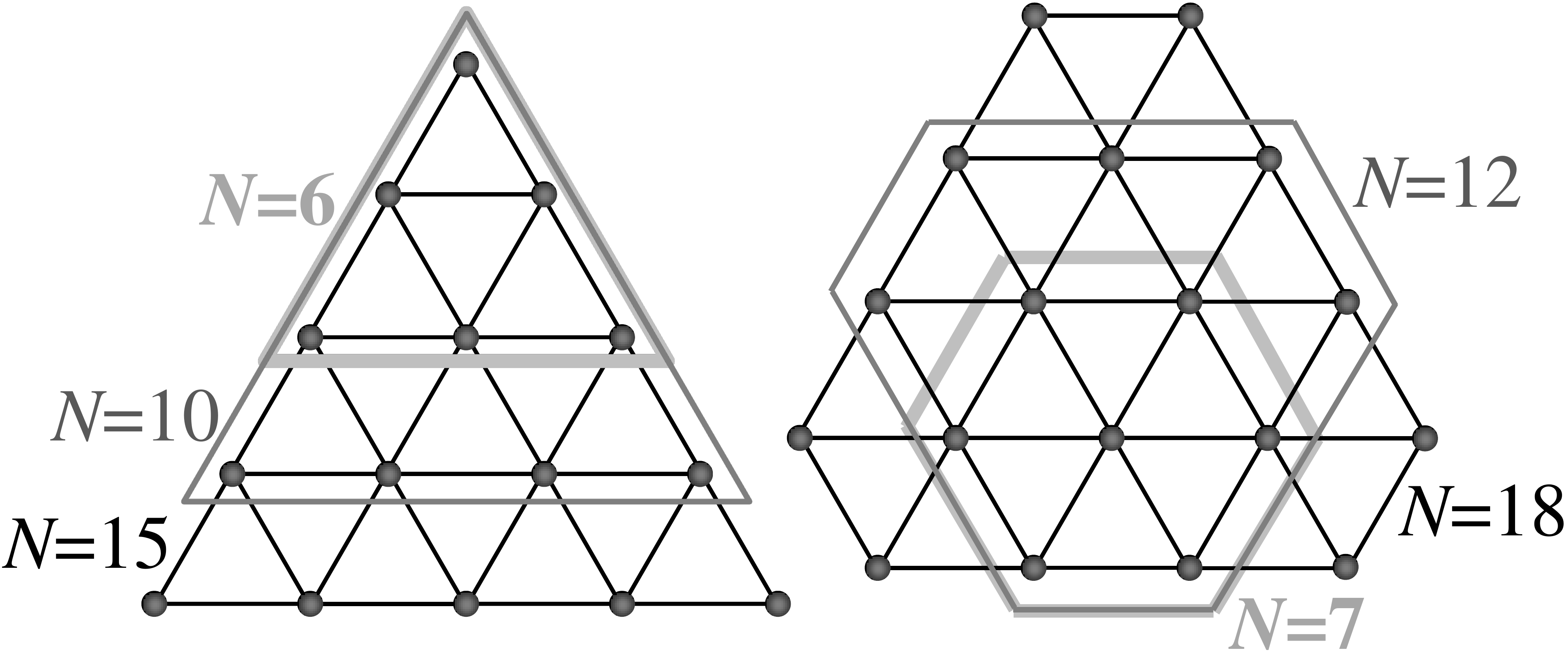}
	\caption{
	  {\bf Chiral correlations from ED.} {\bf Upper left:} Already for small systems ($N=15$), chiral correlations are appreciably smaller in the N\'eel phases than in the rest of the phase diagram. {\bf Upper right:} 
	  Extrapolating to large lattices, chiral LRO apparently only survives in a small region around $(J'/J,J''/J)=(1,1)$, lending support to an extended non-magnetic phase surrounding a spiral phase. 
{\bf Below:} The geometries used in the extrapolation are chosen for symmetry upon rotation by $60^\circ$ and equal number of $J$, $J'$, and $J''$ bonds. 
	  \label{fig:phd_ED_CC}
	}
\end{figure}
As can be seen from the ED results of the $N=15$ lattice (Fig.~\ref{fig:phd_ED_CC}, left panel), the chiral correlations are relatively small in the N\'eel phases and largest in the spiral phase around $(J'/J,J''/J)\approx(1,1)$. However, at this lattice size, there are still appreciable chiral correlations in the rest of the parameter regime. In particular, in the 1D limit, they are only a little smaller than in the spiral phase. Therefore, we also plot in Fig.~\ref{fig:phd_ED_CC}, right panel, an extrapolation to large lattices by $\Psi_{-}(N)=\Psi_{-}(N=\infty)+\frac{c_1}{\sqrt{N}}+\frac{c_2}{N}+\frac{c_3}{N^{3/2}}$, where we use the known form for the leading finite-size behavior \cite{Momoi1994} but also include subleading corrections due to the small systems under consideration (our data comes from lattices with $N=7,10,12,15,18$, all chosen to have the same number of $J$, $J'$, and $J''$ bonds, as sketched at the bottom of Fig.~\ref{fig:phd_ED_CC}). 
This shows a clear trend, namely that the chiral correlations only survive in a small region around $(J'/J,J''/J)=(1,1)$, roughly where the vanishing gap indicated the spiral phase \footnote{The smaller peaks around $(J'/J,J''/J)=(3,1)$ and $(J'/J,J''/J)=(1,3)$ result from the strong geometry dependence for the small lattices used. It can be understood that these peaks are artifacts, because it is highly implausible that the chiral LRO first disappears when increasing the one-dimensionality and then finds a revival.}. This would mean that outside this region there is no spiral LRO. 

With this, we have several independent observations from ED indicating the existence of a magnetically disordered phase surrounding the spiral phase: the increase of the gap when leaving the central region around $(J'/J,J''/J)=(1,1)$ and the disappearance of chiral LRO for large lattices both suggest that there is no spiral LRO in this region. On the other hand, the predominant order is at incommensurate ordering vectors, indicating that this phase is also not N\'eel ordered. Therefore, it seems natural to assume that this region could host a non-magnetic phase, possibly gapped far away from the 1D limit and gapless close to it, consistent with MSWT.

\subsection{MSWT spin-wave dispersion relations}

\begin{figure*}
	\centering
\includegraphics[width=0.8\textwidth]{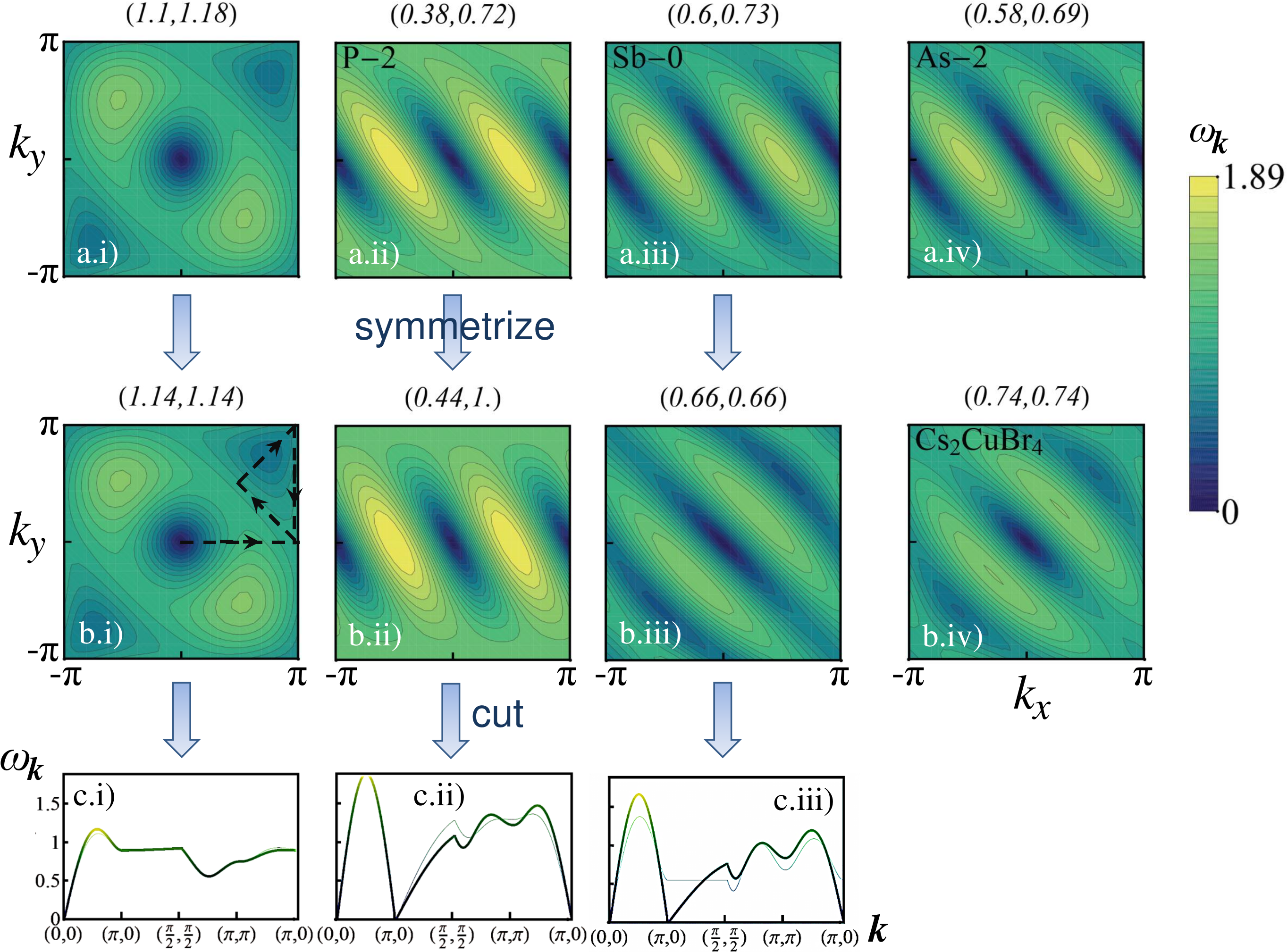}
	\caption{
	  {\bf Spin-wave dispersion from MSWT,} normalized to $1+J'+J''$.
	  \label{fig:phd_MSWT_dispersionrelations}
	  {\bf Top row:} parameters corresponding to a completely anisotropic point in the spiral phase (a.i), and the magnetically ordered materials cited in Ref.~\cite{Scriven2011} (a.ii-iv). 
		{\bf Middle row:} symmetrizing the two closer couplings can change the dispersion relations (b.i-iii). (b.iv): dispersion relation for Cs$_2$CuBr$_4$. 
		{\bf Bottom row:} Cuts along the path indicated in (b.i) allow to more quantitatively compare the completely anisotropic dispersion relation taken from row (a) (thick line) to the corresponding symmetrized one from row (b) (thin line).
	}
\end{figure*}

Finally, to connect to experiment, we provide the spin-wave dispersion relations $\omega_{\vect{k}}$, as computed from MSWT, Eq.~\eqref{disp}. In Fig.~\ref{fig:phd_MSWT_dispersionrelations}, we show parameters corresponding to a point from the spiral phase and the magnetically ordered materials listed in Table~\ref{tab:materials}. 
We also provide (where applicable) a comparison to the dispersion relation which would result if two of the couplings were equal. These comparisons can be seen more quantitatively in the cuts (c.i-iii) shown in the lowest row of Fig.~\ref{fig:phd_MSWT_dispersionrelations}. 
For the point from the spiral phase (a.i), the symmetrization (b.i) does not significantly change the dispersion relation, but for P-2 and, especially, for Sb-0, the differences are considerable. The latter in particular changes even qualitatively since a symmetrization would put it instead of into a N\'eel phase into a spiral phase. These differences seem significant enough to be measurable in experiment. Such a measurement could allow to quantify the actual magnitude of coupling anisotropies. 

\section{Conclusion\label{cha:conclusion}}

In conclusion, we have provided a thorough analysis of the ground-state phase diagram of the quantum Heisenberg SCATL. 
Using various observables from modified spin-wave theory supplemented with ordering-vector optimization, and supported by exact diagonalization data, we have found that quantum fluctuations stabilize N\'eel order with respect to the classical phase diagram. Further, they reduce the extent of the spiral phase, which seems to be entirely surrounded by a quantum disordered region. 
This result, which constitutes our first main finding, is supported by the breakdown of MSWT, together with the strong decrease of the order parameter and the spin stiffness. 
While MSWT cannot be applied to studying this region, the fact that \emph{no} semiclassical reference state describable by an ordering vector yields a stable solution is highly suggestive of a magnetically-disordered phase of purely quantum origin. Hence, our results outline a very promising candidate region for such exotic states, meriting further research with more sophisticated theoretical methods or experimental setups. 

The possible existence of quantum-disordered phases is further corroborated by ED data, where a finite gap and a vanishing chiral correlation make spiral LRO seem unlikely, while the location of the structure-factor peak at incommensurate wave vectors seems to preclude N\'eel LRO. Also, the strong decrease of the ED structure-factor peak appears to support this interpretation. 
Further, we found some indications that very close to the 1D limit the transition from the N\'eel phase to the putative disordered region could be qualitatively different from what happens at larger $\alpha$. 

A complete encircling of the spiral phase by disordered phases could naturally explain the succession of a gapped and a gapless non-magnetic phase at the low-$\alpha$ limit of the SATL. The gapless quantum-disordered phase would be continuously connected to the limit of decoupled chains, while the additional anisotropy of the SCATL would adiabatically connect the gapped quantum-disordered phases at small and large $\alpha$. 
Therefore, the additional anisotropy has great potential to deliver new insights into the nature of these kind of phases. 

Our second main finding is connected to experimental results: measurements find magnetic LRO in materials which theoretical analyses on the SATL predict to be magnetically disordered. We show that this discrepancy finds a simple explanation in the additional anisotropy of the SCATL, which is neglected in the SATL. Taking it into account, we predict these material to lie in magnetically ordered phases, in accordance to experiment. These findings show the importance of the complete lattice anisotropy for the explanation of recent experiments. 

Finally, we provided spin-wave dispersion relations, a comparison to which might allow to probe the additional anisotropy experimentally.\\

\textbf{Acknowledgments}
I gratefully acknowledge fruitful discussions with Tommaso Roscilde, Roman Schmied, and Luca Tagliacozzo. Also, I would like to thank Ben Powell for drawing my attention to the SCATL model. 
This work has been supported by the Catalunya Caixa, 
Spanish MICINN (FIS2008-00784), AAII-Hubbard, EU Project AQUTE, 
the Austrian Science Fund through SFB F40 FOQUS, the DARPA OLE program, 
and ERC Grant QUAGATUA.

\appendix

\section{MSWT formalism\label{cha:appendixMSWT}}

In this Appendix, we shortly review the MSWT for Heisenberg
antiferromagnets (see~\cite{Hauke2011}; for a full description of the approach -- as applied to XY models -- 
see~\cite{Hauke2010}).

A fundamental assumption of spin-wave theory is that the ground state has LRO with ordering vector $\vect{Q}$. Hence, it is convenient to rotate the local reference system as
\begin{subequations}
\label{rot}
\begin{eqnarray}
S_i^{\,x}  &=& - \sin\left(\vect{Q}\cdot\vect{r}_i\right) S_i^{\,\eta} + \cos\left(\vect{Q}\cdot\vect{r}_i\right) S_i^{\,\zeta}\,,\\
    S_i^{\,y} &=& \phantom{-}\cos\left(\vect{Q}\cdot\vect{r}_i\right) S_i^{\,\eta} + \sin\left(\vect{Q}\cdot\vect{r}_i\right) S_i^{\,\zeta}\,,\\
    S_i^{\,z} &=& - S_i^{\,\xi}\,.
\end{eqnarray}
\end{subequations}
Then $S_i^{\,\zeta}$, which will be the quantization axis, lies parallel to the classical spin $\vect{S}_i=\left(\cos\left(\vect{Q}\cdot\vect{r}_i\right),\sin\left(\vect{Q}\cdot\vect{r}_i\right),0\right)$. This defines the classical reference state. 
We do not make any assumption on the ordering vector $\bm Q$. In particular, it may well differ from the one of the classical limit ($\vect{Q}^{\rm cl}$). 

Spin waves around this reference state can be described by the Dyson--Maleev (DM) transformation \cite{Dyson1956,Maleev1957}, which maps the physical spins to interacting bosons, 
\begin{subequations}
\label{DM}
\begin{eqnarray}
    S_i^{\,-} &\to& \frac{1}{\sqrt{2 S}}\left(2 S - a_i^\dagger a_i\right)a_i\,,\\
    S_i^{\,+} &\to& \sqrt{2 S}\, a_i^\dagger,\\
    S_i^{\,\zeta} &\to& -S+a_i^\dagger a_i\,,
\end{eqnarray}
\end{subequations}
where $S_i^{\,\pm}\equiv S_i^{\,\xi}\pm i S_i^{\,\eta}$. 
 
\begin{widetext}
This transformation maps the Hamiltonian, Eq.~\eqref{eq:HS}, to the non-linear bosonic Hamiltonian
\begin{eqnarray}
\label{H4}
\null\hspace*{-2cm}
    {\cal H}&=&\frac 1 4 \sum_{\braket{i,j}} J_{ij} \left\lbrace \phantom{+4} \left[ 2 S \left( a_i^\dagger a_j + a_i a_j^\dagger \right) - a_i^\dagger a_j^\dagger a_j a_j - a_i^\dagger a_i a_i a_j^\dagger \right] \left(1+\cos\left(\vect{Q}\cdot\vect{r}_{ij}\right)\right) \right. \nonumber \\
 \null\hspace*{-2cm}   & &\phantom{\frac 1 4 \sum_{\braket{i,j}} J_{ij} \lbrace 4} + \left[ 2 S \left( a_i^\dagger a_j^\dagger + a_i a_j \right) - a_i a_j^\dagger a_j a_j - a_i^\dagger a_i a_i a_j \right] \left(1-\cos\left(\vect{Q}\cdot\vect{r}_{ij}\right)\right) \\
 \null\hspace*{-2cm}   & &\phantom{\frac 1 4 \sum_{\braket{i,j}} J_{ij} \lbrace 4} \left.+\,4 \left[ S^2 - S\left(a_i^\dagger a_i + a_j^\dagger a_j\right)+a_i^\dagger a_i a_j^\dagger a_j\right] \cos\left(\vect{Q}\cdot\vect{r}_{ij}\right) + {\cal O}\left( \frac{1}{S}\right) \quad \right\rbrace \,, \nonumber
\end{eqnarray}
\end{widetext}
where $a_i$ ($a_i^{\dagger}$) destroys (creates) a DM boson at site $i$, and $S$ is the length of the spin.
Here, we neglected the kinematic constraint which restricts the DM-boson density $n$ to the physical subspace $n < 2S$. Moreover, we dropped terms with six boson operators, which are of order  ${\cal O}[n/(2S)^3]$ and are negligible for $n/(2S)<1$. 
Using Wick's theorem \cite{Fetter1971}, and defining the correlators $\braket{a_i^\dagger a_j}=F\left( \vect{r}_{ij} \right)-\frac 1 2 \delta_{ij}$ and $\braket{a_i a_j}= \braket{a_i^\dagger a_j^\dagger} \,\, = \,\, G\left( \vect{r}_{ij} \right)$, the expectation value $E\equiv\braket{\cal H}$ can be written as
\begin{eqnarray}
    E &= & \frac 1 2 \sum_{\braket{i,j}} J_{ij} \left\lbrace \right.  \\ 
    & & \left[S+\frac 1 2 - F\left( 0 \right) + F\left( \vect{r}_{ij} \right) \right]^2 \left(1+\cos\left(\vect{Q}\cdot\vect{r}_{ij}\right)\right) \nonumber \\  
    &- &\left. \left[S+\frac 1 2 - F\left( 0 \right) + G\left( \vect{r}_{ij} \right) \right]^2 \left(1-\cos\left(\vect{Q}\cdot\vect{r}_{ij}\right)\right)\,\right\rbrace \,. \nonumber
\end{eqnarray}
After Fourier transforming, $a_{\vect{k}}=\frac{1}{\sqrt{N}}\sum_i a_i\, \ue^{-i \vect{k}\cdot\vect{r}_i}$, and a subsequent Bogoliubov transformation, $\alpha_{\vect{k}\phantom{-}} = \phantom{-}\cosh\theta_{\vect{k}}\, a_{\vect{k}} - \sinh\theta_{\vect{k}} \, a_{-\vect{k}}^\dagger$, and $\alpha_{-\vect{k}}^\dagger = -\sinh\theta_{\vect{k}} \, a_{\vect{k}} + \cosh\theta_{\vect{k}} \, a_{-\vect{k}}^\dagger$, we 
minimize the free energy $\mathcal{F}$ under the constraint of vanishing magnetization at each site, $\braket{a_i^{\dagger} a_i} = S$, which is known as Takahashi's modification \cite{Takahashi1989}. 
This yields a set of self-consistent equations, 
\begin{equation}
    \label{tanh2th}
    \tanh 2\theta_{\vect{k}}=\frac{A_{\vect{k}}}{B_{\vect{k}}}
\end{equation}
with
\begin{subequations}
    \label{AkBk}
\begin{eqnarray}
    \label{Ak}
    A_{\vect{k}} & = & \frac 1 N \sum_{\braket{i,j}} J_{ij} \left(1 - \cos\left({\vect{Q}\cdot\vect{r}_{ij}}\right)\right) G_{ij} \,\ue^{i\vect{k}\cdot\vect{r}_{ij}}\,,\\
    \label{Bk}
    B_{\vect{k}} &= &\frac 1 N \sum_{\braket{i,j}} J_{ij} \left[ \left(1 - \cos\left({\vect{Q}\cdot\vect{r}_{ij}}\right)\right) G_{ij} \right. \\
    & & \left. - \left(1 + \cos\left({\vect{Q}\cdot\vect{r}_{ij}}\right)\right) F_{ij} \left(1-\ue^{i\vect{k}\cdot\vect{r}_{ij}}\right)\right] -\mu \nonumber \,,
\end{eqnarray}
\end{subequations}
where $\mu$ is the Lagrange multiplier for Takahashi's constraint.
The spin-wave spectrum reads
\begin{equation}
\label{disp}
\omega_{\vect{k}}=\sqrt{B_{\vect{k}}^2-A_{\vect{k}}^2}\,.
\end{equation}
At $T=0$, $\mu$ vanishes, which implies the disappearance of the gap at $\vect{k}=0$ that may exist for finite temperature. This is a necessary condition for magnetic LRO, and enables Bose condensation in the $\vect{k}=0$ mode. 
Separating out its contribution, 
\eq{
\braket{a_{\vect{k}=0}^{\dagger} a_{\vect{k}=0}}/N=\braket{a_{\vect{k}=0} a_{\vect{k}=0}}/N\equiv M\,,
\label{eq:orderParameterMSWT}
}
which corresponds to the magnetic order parameter, 
one arrives at the zero-temperature equations
\begin{subequations}
\label{FG}
\begin{eqnarray}
    F_{ij}&=&M + \frac 1 {2 N} \sum_{\vect{k}\neq 0} \frac{B_{\vect{k}}} {\omega_{\vect{k}}}\cos\left(\vect{k}\cdot\vect{r}_{ij}\right)\label{Fij}\,,\\
    G_{ij}&=&M + \frac 1 {2 N} \sum_{\vect{k}\neq 0} \frac{A_{\vect{k}}} {\omega_{\vect{k}}}\cos\left(\vect{k}\cdot\vect{r}_{ij}\right)\label{Gij}\,,
\end{eqnarray}
\end{subequations}
and the constraint of vanishing magnetization at each site becomes
\begin{equation}
    \label{constr2}
    S+\frac 1 2 = M + \frac 1 {2 N} \sum_{\vect{k}\neq 0} \frac{B_{\vect{k}}} {\omega_{\vect{k}}}.
\end{equation}

It is not \emph{a priori} clear that the classical ordering vector $\vect{Q}^{\mathrm{cl}}$ correctly describes the LRO in the quantum system. 
To account for a competition between LRO at different ordering vectors $\vect{Q}$, we extend the MSWT procedure by optimizing the free energy $\mathcal{F}$ with respect to the ordering vector $\vect{Q}$.
This yields two additional equations which must be added to the set of self-consistent equations, 
\begin{subequations}
\label{Qs}
\begin{equation}
    \label{Qx}
    \frac\partial{\partial Q_x} \mathcal{F}=-\frac 1 2 \sum_{\braket{i,j}} J_{ij} \sin\left(\vect{Q}\cdot\vect{r}_{ij}\right)r_{ij}^x\left[F_{ij}^2+G_{ij}^2\right] = 0 \,,
\end{equation}
\begin{equation}
    \label{Qy}
    \frac\partial{\partial Q_y} \mathcal{F}=-\frac 1 2 \sum_{\braket{i,j}} J_{ij} \sin\left(\vect{Q}\cdot\vect{r}_{ij}\right)r_{ij}^y\left[F_{ij}^2+G_{ij}^2\right]=0\,.
\end{equation}
\end{subequations}

The values of $F_{ij}$ and $G_{ij}$ can now be calculated by solving self-consistently Eqs.~(\ref{AkBk}--\ref{Qs}).
Through Wick's theorem the knowledge of the quantities $F_{ij}$ and $G_{ij}$ 
determines the expectation value of any observable.

\end{document}